\documentclass[a4paper,11pt]{article}
\pdfoutput=1 

\usepackage{jheppub} 

\usepackage[T1]{fontenc} 

\usepackage{amsthm}

\usepackage{slashed} 

\title{\boldmath R-symmetries and curvature constraints in A-twisted heterotic Landau-Ginzburg models}

\author{Richard S. Garavuso}

\affiliation{Physical Sciences, Kingsborough Community College, The City University of New York,\\ 2001 Oriental Boulevard, Brooklyn, NY 11235-2398, USA}

\emailAdd{richard.garavuso@kbcc.cuny.edu}
 
\abstract{In this paper, we discuss various aspects of a class of A-twisted heterotic Landau-Ginzburg models on a K\"{a}hler variety
$ X $.
We provide a classification of the R-symmetries in these models which allow the 
A-twist to be implemented, focusing on the case in which the gauge bundle is either a deformation of the tangent bundle of
$ X $
or a deformation of a sub-bundle of the tangent bundle of
$ X $.
Some anomaly-free examples are provided.
The curvature constraint imposed by supersymmetry in these models when the superpotential is not holomorphic is reviewed.
Constraints of this nature have been used to establish properties of analogues of pullbacks of Mathai-Quillen forms which arise in the correlation functions of the corresponding A-twisted or B-twisted heterotic Landau-Ginzburg models.
The analogue most relevant to this paper is a deformation of the pullback of a Mathai-Quillen form.
We discuss how this deformation may arise in the class of models studied in this paper.  
We then comment on how analogues of pullbacks of Mathai-Quillen forms not discussed in previous work may be obtained.
Standard Mathai-Quillen formalism is reviewed in an appendix.
We also include an appendix which discusses the deformation of the pullback of a Mathai-Quillen form.
}

\begin{document} 
\maketitle
\flushbottom
\numberwithin{equation}{section}
\numberwithin{figure}{section}
\numberwithin{table}{section}
\newtheorem{theor}{Theorem}[section]
\newtheorem{defi}[theor]{Definition}
\newtheorem{prop}[theor]{Proposition}
\newtheorem{corol}[theor]{Corollary}
\newtheorem{exa}[theor]{Example}
\newtheorem{remar}[theor]{Remark}
\section{\label{section:Introduction}Introduction}

A Landau-Ginzburg model is a nonlinear sigma model with a superpotential.
For a \emph{heterotic} Landau-Ginzburg model
\cite{Witten:Phases, DistlerKachru:0-2-Landau-Ginzburg, AdamsBasuSethi:0-2-Duality, MelnikovSethi:Half-twisted, GuffinSharpe:A-twistedheterotic, MelnikovSethiSharpe:Recent-Developments, GaravusoSharpe:Analogues}, 
the nonlinear sigma model possesses only
$ (0, 2) $
supersymmetry and the superpotential is a Grassmann-odd function of the superfields which may or may not be holomorphic.

Heterotic Landau-Ginzburg models have field content consisting of
$ (0, 2) $ 
bosonic chiral superfields
\[ 
\Phi^i = (\phi^i, \psi^i_+ )
\]
and
$ (0, 2) $ 
fermionic chiral superfields 
\[
\Lambda^a 
  = \left( \lambda^a_-, 
           H^a, 
           E^a
    \right) ,
\]
along with their conjugate antichiral superfields
\[ 
\Phi^{\overline{\imath}} 
  = \left(
           \phi^{\overline{\imath}}, 
           \psi^{\overline{\imath}}_+ 
    \right)
\]
and
\[ 
\Lambda^{\overline{a}} 
  = \left( 
           \lambda^{\overline{a}}_-, 
           \overline{H}^{\overline{a}}, 
           \overline{E}^{\overline{a}}
    \right) .
\]
The 
$ \phi^i $ 
are local complex coordinates on a K{\"a}hler variety 
$ X $. 
The 
$ E^a $
are local smooth sections of a Hermitian vector bundle 
$ \mathcal{E} $
over 
$ X $,
i.e. 
$ E^a \in \Gamma(X, \mathcal{E}) $. 
The
$ H^a $
are nonpropagating auxiliary fields.
The fermions couple to bundles as follows:
\[
\begin{aligned}
\psi^i_+ 
  \in \Gamma
      \left(
             K^{1/2}_{\Sigma}
             \otimes
             \Phi^* \!
             \left(
                    T^{1,0} 
                    X
             \right)      
      \right),
\qquad
&
\lambda^a_- 
  \in \Gamma
      \left(
             \overline{K}^{1/2}_{\Sigma}
             \otimes
             \left(
                    \Phi^* 
                    \overline{\mathcal{E}}
             \right)^{\vee}
      \right),
\\
\psi^{\overline{\imath}}_+
  \in \Gamma
      \left(
             K^{1/2}_{\Sigma}
             \otimes
             \left(
                    \Phi^* \!
                    \left(
                           T^{1,0}
                           X
                    \right)       
             \right)^{\vee}
      \right),
\qquad
&
\lambda^{\overline{a}}_-
  \in \Gamma
      \left(
             \overline{K}^{1/2}_{\Sigma}
             \otimes
             \Phi^* 
             \overline{\mathcal{E}}
      \right),                    
\end{aligned}
\]
where
$ \Phi: \Sigma \rightarrow X $
and
$ K_{\Sigma} $
is the canonical bundle on the worldsheet
$ \Sigma $.

In 
\cite{GuffinSharpe:A-twistedheterotic},
heterotic Landau-Ginzburg models with superpotential of the form
\begin{equation}
  W = \Lambda^a \,
      F_a \, ,
\label{superpotential}      
\end{equation}      
where 
$ 
  F_a 
  \in
  \Gamma
  \left(
         X, 
         \mathcal{E}^{\vee}  
  \right)
$
were considered.
It was claimed in 
\cite{GaravusoSharpe:Analogues}
that, when the superpotential 
\eqref{superpotential}
is not holomorphic, supersymmetry imposes a constraint which relates the nonholomorphic parameters of the superpotential to the Hermitian curvature.
The details supporting that claim were worked out in
\cite{Garavuso:Curvature, Garavuso:Nonholomorphic}
for the case
$ E^a \equiv 0 $.
This curvature constraint has been used in
\cite{GaravusoSharpe:Analogues}
to establish properties of 
analogues of pullbacks of Mathai-Quillen forms.
These analogues arise in the correlation functions of the corresponding A-twisted or B-twisted heterotic Landau-Ginzburg models.

In this paper, we will study certain aspects of A-twisted heterotic Landau-Ginzburg models with superpotential
\eqref{superpotential}
and
$ E^a \equiv 0 $.
Such models yield the A-twisted
$ (2,2) $ 
Landau-Ginzburg models of
\cite{GuffinSharpe:A-twisted}
when
$ \mathcal{E} = TX $ 
and
$ \Lambda^i \, F_i = \Lambda^i \, \partial_i W^{(2,2)} $,
where
$ W^{(2,2)} $
is the (2,2) superpotential.
Although R-symmetries for (2,2) Landau-Ginzburg models have been classified, this has not been done for heterotic Landau-Ginzburg models.
Furthermore, for (2,2) Landau-Ginzburg models, a classification has been given only for the case of holomorphic superpotentials \cite{KachruWitten:Computing}.
We will provide a classification of the R-symmetries which allow the A-twist to be implemented, focusing on the case in which
$ \mathcal{E} $
is either a deformation of 
$ TX $
or a deformation of a sub-bundle of
$ TX $.
The curvature constraint imposed by supersymmetry in these models when the superpotential is not holomorphic will be reviewed.
The corresponding analogue of the pullback of a Mathai-Quillen form is a deformation of the pullback of a Mathai-Quillen form.
We will discuss how this deformation may arise in the class of models studied in this paper.
We will then comment on how analogues of pullbacks of Mathai-Quillen forms not discussed in previous work may be obtained.

This paper is organized as follows: 
The A-twist will be discussed in section 
\ref{section:A-twist}.
A classification of the corresponding R-symmetries, along with some anomaly-free examples, will be given in section
\ref{section:R-symmetries}.
The curvature constraint imposed by supersymmetry when the superpotential is not holomorphic will be reviewed in section
\ref{section:Curvature}.
In section
\ref{section:Physical}, we will discuss how an analogue of a pullback of a Mathai-Quillen form may arise in the class of heterotic Landau-Ginzburg models discussed in this paper.
In section
\ref{section:Summary},
we will summarize our results and comment on how analogues of pullbacks of Mathai-Quillen forms not discussed in previous work may be obtained.
Appendix
\ref{section:Review-MQ-formalism}
will review standard Mathai-Quillen formalism
\cite{MathaiQuillen:Superconnections,
BerlinGetzlerVergne:Heat,
Kalkman:BRST,
Blau:The-Mathai-Quillen,
Wu:On-the-Mathai-Quillen,
CordesMooreRamgoolam:Lectures,
Wu:Mathai-Quillen}.
Finally, appendix
\ref{section:Deformation-pullback-Mathai-Quillen}
will discuss the analogue that is most relevant to this paper, i.e. a deformation of the pullback of a Mathai-Quillen form.
\section{\label{section:A-twist}A-twist}

Let
$ X $ 
be a K{\"a}hler variety with metric
$ g $, 
antisymmetric tensor
$ B $,
local real coordinates
$ \phi^{\mu} $, 
and local complex coordinates 
$ \phi^i $
with complex conjugates
$ \phi^{\overline{\imath}} $.
Furthermore, let 
$ \mathcal{E} $
be a vector bundle over
$ X $
with Hermitian fiber metric
$ h $.
We consider the action
\cite{GuffinSharpe:A-twistedheterotic}
of an A-twisted heterotic Landau-Ginzburg model on
$ X $ 
with gauge bundle 
$ \mathcal{E} $:

\begin{align}
\label{action}
S &= 2t \int_{\Sigma} d^2 z 
     \left[ 
            \frac{1}{2} 
            \left( 
                   g_{\mu \nu} + i B_{\mu \nu}
            \right)
            \partial_z 
            \phi^{\mu} 
            \partial_{\overline{z}} 
            \phi^{\nu}
          + i 
            g_{\overline{\imath} i}
            \psi_+^{\overline{\imath}}
            \overline{D}_{\overline{z}} \psi_+^i
          + i 
            h_{a \overline{a}} 
            \lambda_-^a
            D_z \lambda_-^{\overline{a}}
     \right. 
\nonumber
\\
  &\phantom{= 2t \int_{\Sigma} d^2 z \left[ \right.}
   +
     \biggl. 
     F_{i \overline{\imath} a \overline{a}} \,
     \psi_+^i
     \psi_+^{\overline{\imath}}
     \lambda_-^a
     \lambda_-^{\overline{a}}
   + h^{a \overline{a}} 
     F_{a} 
     \overline{F}_{\overline{a}}
   + \psi_+^i 
     \lambda_-^{a} 
     D_i F_a 
   + \psi_+^{\overline{\imath}}
     \lambda_-^{\overline{a}} 
     \overline{D}_{\overline{\imath}} \overline{F}_{\overline{a}}
     \biggr].
\end{align}
Here,
$ t $ 
is a coupling constant,
$ \Sigma $
is a Riemann surface,
$ d^2 z = -i \, dz \wedge d{\overline{z}} $, 
$ 
  F_a  
  \in 
  \Gamma
  \left(
         X, 
         \mathcal{E}^{\vee}
  \right) 
$,
and
\[
\begin{aligned}
\overline{D}_{\overline{z}} \,
\psi^i_+
   &= \overline{\partial}_{\overline{z}} \,
      \psi_+^i
    + \overline{\partial}_{\overline{z}} \,
      \phi^j \,
      \Gamma^i_{jk}
      \psi^k_+ \, ,
&D_z \lambda^{\overline{a}}_-
  &= \partial_z 
     \lambda_-^{\overline{a}}
   + \partial_z \phi^{\overline{\imath}}
     A^{\overline{a}}_{\overline{\imath} \overline{b}} \,
     \lambda^{\overline{b}}_- \, ,
\\
D_i F_a
  &= \partial_i F_a
   - A^{b}_{i a}
     F_b \, ,
& \overline{D}_{\overline{\imath}}
  \overline{F}_{\overline{a}}
  &= \overline{\partial}_{\overline{\imath}} \,
     \overline{F}_{\overline{a}}
   - A^{\overline{b}}_{\overline{\imath} \, \overline{a}} \,
     \overline{F}_{\overline{b}} \, ,
\\
A^b_{i a} 
  &= h^{b \overline{b}} \,
     h_{\overline{b} a, i} \, ,
& A^{\overline{b}}_{\overline{\imath} \, \overline{a}} 
  &= h^{\overline{b} b} \, 
     h_{b \overline{a}, \overline{\imath}} \, ,     
\\
\Gamma^i_{jk}
  &= g^{i \overline{\imath}} \,
     g_{\overline{\imath} k, j} \, ,         
& F_{i \overline{\imath} a \overline{a}}  
   &= h_{a \overline{b}} \,
      A^{\overline{b}}_{\overline{\imath} \, \overline{a}, i} \, .
\end{aligned}
\]
The A-twist is defined by choosing the fermions couple to bundles as follows:
\[
\begin{aligned}
\psi^i_+ 
  \in \Gamma
      \left(
             \Phi^* \!
             \left(
                    T^{1,0} 
                    X
             \right)      
      \right),
\qquad
&
\lambda^a_- 
  \in \Gamma
      \left(
             \overline{K}_{\Sigma}
             \otimes
             \left(
                    \Phi^* 
                    \overline{\mathcal{E}}
             \right)^{\vee}
      \right),
\\
\psi^{\overline{\imath}}_+
  \in \Gamma
      \left(
             K_{\Sigma}
             \otimes
             \left(
                    \Phi^* \!
                    \left(
                           T^{1,0}
                           X
                    \right)       
             \right)^{\vee}
      \right),
\qquad
&
\lambda^{\overline{a}}_-
  \in \Gamma
      \left(
             \Phi^* 
             \overline{\mathcal{E}}
      \right),                    
\end{aligned}
\label{A-twist-fermions}
\]
where
$ \Phi: \Sigma \rightarrow X $
and
$ K_{\Sigma} $
is the canonical bundle on
$ \Sigma $.
Anomaly cancellation requires
\cite{GuffinSharpe:A-twistedheterotic,KatzSharpe:Notes,Sharpe:Notes}
\begin{equation}
\label{anomaly-cancellation-conditions}
\Lambda^{\textrm{top}} 
\mathcal{E}^{\vee}
  \simeq K_X \, ,
\qquad
\textrm{ch}_2
\left(
       \mathcal{E}
\right)
  = \textrm{ch}_2
    \left(
           TX
    \right) .         
\end{equation}

The action 
\eqref{action}
is invariant on-shell under the supersymmetry transformations
\begin{equation}
\begin{aligned}
\delta \phi^i
  &= i 
     \alpha_-
     \psi^i_+ \, ,
\\
\delta \phi^{\overline{\imath}}
  &= 0 \, ,          
\\
\delta \psi^i_+ 
  &= 0 \, ,
\\
\delta \psi^{\overline{\imath}}_+
  &=
   - \alpha_-
     \partial_z \phi^{\overline{\imath}} \, ,
\\
\delta \lambda^{a}_-
  &= 
   - i
     \alpha_-
     \psi^j_+ \,
     A^a_{j b} \,
     \lambda^{b}_- 
   + i
     \alpha_-
     h^{a \overline{a}} \,
     \overline{F}_{\overline{a}} \, ,
\\
\delta \lambda^{\overline{a}}_-
  &= 0                 
\end{aligned}
\label{SUSY}
\end{equation}
up to a total derivative.
Since we have integrated out the auxiliary fields
$ H^a $,
one may use the 
$ \lambda^a_- $ 
equation of motion
\begin{equation}
\label{eom:lambda-super-a-sub-minus}
\lambda^a_- :
\quad
  i
  h_{a \overline{a}}
  D_z \lambda^{\overline{a}}_-
+ F_{i \overline{\imath} a \overline{a}} \,
  \psi^i_+
  \psi^{\overline{\imath}}_+
  \lambda^{\overline{a}}_-
- \psi^i_+
  D_i
  F_a
= 0 \, ,  
\end{equation}
to show 
\cite{Garavuso:Curvature}
that the action
\eqref{action}
can be written
\begin{equation}
\label{action:Q-exact+non-exact}
S = it \int_{\Sigma} d^2 z \,
    \left\{
            Q, V
    \right\}
  + t \int_{\Sigma}
    \Phi^*(K)
  + 2 t 
    \int_{\Sigma} d^2 z
    \left(
           \psi_+^{\overline{\imath}}
           \lambda_-^{\overline{a}} 
           \overline{D}_{\overline{\imath}} \overline{F}_{\overline{a}}
         - \psi_+^i 
           \lambda_-^{a} 
           D_a F_a
    \right),               
\end{equation}
where
\[
\label{BRST}
\begin{aligned}
\left\{ 
        Q, \phi^i 
\right\}
  &= - \psi^i_+ \, ,
\qquad
&\left\{ 
         Q, \phi^{\overline{\imath}} 
 \right\} &= 0 \, ,
\\
\left\{ 
        Q, \psi^i_+ 
\right\} 
  &= 0 \, ,
\qquad
&\left\{ 
         Q, \psi^{\overline{\imath}}_+ 
 \right\} 
  &= - i \partial_z \phi^{\overline{\imath}} \, ,
\\
\left\{ 
        Q, \lambda^a_- 
\right\}
  &= \psi^j_+ 
     A^{a}_{j b} \lambda^b_- 
   - h^{a \overline{a}} \,
     \overline{F}_{\overline{a}} \, ,
\qquad
&\left\{ 
         Q, \lambda^{\overline{a}}_- 
 \right\}
  &= 0  
\end{aligned}
\]
are the BRST transformations
($ \delta f = - i \alpha_- \{ Q,f \} $, where $ f $ is any field),
\[
V = 2 \left(
               g_{\overline{\imath} i}
               \psi^{\overline{\imath}}_+ 
               \overline{\partial}_{\overline{z}} \phi^i
             + i 
               \lambda^a_- 
               F_a         
      \right),
\]
and 
\[
\int_{\Sigma} \Phi^*(K) 
   = \int_{\Sigma} d^2 z 
     \left(
            g_{i \overline{\imath}}
          + i 
            B_{i \overline{\imath}}  
     \right)
     \left(
            \partial_z \phi^i \,
            \overline{\partial}_{\overline{z}} \phi^{\overline{\imath}}
          - \overline{\partial}_{\overline{z}} \phi^i
            \partial_z \phi^{\overline{\imath}} 
     \right)          
\]
is the integral over the worldsheet 
$ \Sigma $
of the
pullback to
$ \Sigma $
of the complexified K{\"a}hler form
\[
K = 
  - i 
    \left(
           g_{i \overline{\imath}}
         + i
           B_{i \overline{\imath}} 
    \right)       
    d \phi^i
    \wedge
    d \phi^{\overline{\imath}} \, .
\]
\section{\label{section:R-symmetries}R-symmetries}

Let us now discuss the R-symmetries which allow the A-twist described in section
\ref{section:A-twist}
to be obtained.
A classification of these R-symmetries will be given in section
\ref{subsection:Classification}.
Some anomaly-free examples will be given in section
\ref{subsection:Examples}.
\subsection{\label{subsection:Classification}Classification}

For 
$ F_a \equiv 0 $, 
the twisting is achieved by tensoring the fields with
\[
K^{- Q_R / 2}_{\Sigma} 
\otimes
\overline{K}^{\, Q_L / 2}_{\Sigma} \, ,
\]
where the fields have charges
$ Q_L $
and
$ Q_R $,
given in table
\ref{charges},
\begin{table}
\begin{center}
\begin{tabular}{|c|c|c|}
\hline
Field &
$ Q_L $ &
$ Q_R $
\\
\hline
$ \phi^i $ &
0 &
0
\\
\hline
$ \phi^{\overline{\imath}} $ &
0 &
0
\\
\hline
$ \psi^i_+ $ &
0 &
1
\\
\hline
$ \psi^{\overline{\imath}}_+ $ &
0 &
$ -1 $
\\
\hline
$ \lambda^a_- $ &
1 &
0
\\
\hline
$ \lambda^{\overline{a}}_- $ &
$ -1 $ &
0
\\
\hline  
\end{tabular}
\caption{Charges when $ F_a \equiv 0 $.}
\label{charges}
\end{center}
\end{table}
\noindent
under
$ U(1)_L $
and
$ U(1)_R $
R-symmetries, respectively.
These R-symmetries defined by 
$ Q_L $ 
and
$ Q_R $
are broken when 
$ F_a \not\equiv 0 $.

Let us consider an
$ F_a $
of the form
\begin{equation}
F_a 
  = \partial_a G 
  + G_a \, .
\end{equation}
Here,
$ G $
is quasihomogeneous and meromorphic, i.e.
\begin{equation}
G
\left(
       \lambda^{n_i}
       \phi^i,
       \lambda^{m_{\overline{\imath}}}
       \phi^{\overline{\imath}} 
\right)
  = \lambda^d \,
    G
    \left(
           \phi^i,
           \phi^{\overline{\imath}} 
    \right),     
\end{equation}
where
$ \lambda \in \mathbf{C}^{\times} $,
$ n_i = - m_{\overline{\imath}} $ 
and
$ d $
are integers, and the deformation
$ G_a $
is chosen to be
\begin{equation}
G_a = \partial_a
      \left[
            \frac{1}{d}
             \sum_i
             n_i
             \left(       
                    \phi^i
             \right)^{\frac{d}{n_i}}        
      \right] 
    = \left(\phi^a\right)^{\frac{d}{n_a} - 1} \, .
\end{equation}
For an
$ F_a $
of this form, we can define new charges
$ Q^{\prime}_{L} = Q_L - Q $
and
$ Q^{\prime}_{R} = Q_R - Q $,
given in table
\ref{new-charges},
\begin{table}
\begin{center}
\begin{tabular}{|c|c|c|c|c|c|}
\hline
Field &
$ Q_L $ &
$ Q_R $ &
$ Q $ &
$ Q^{\prime}_L = Q_L - Q $ &
$ Q^{\prime}_R = Q_R - Q $
\\
\hline
$ \phi^i $ &
0 &
0 & 
$ \alpha_i $ &
$ - \alpha_i $ &
$ - \alpha_i $
\\
\hline
$ \phi^{\overline{\imath}} $ &
0 &
0 &
$ - \alpha_i $ &
$ \alpha_i $ &
$ \alpha_i $
\\
\hline
$ \psi^i_+ $ &
0 &
1 &
$ \alpha_i $ &
$ - \alpha_i $ &
$ 1 - \alpha_i $ 
\\
\hline
$ \psi^{\overline{\imath}}_+ $ &
0 &
$ - 1 $ & 
$ - \alpha_i $ &
$ \alpha_i $ &
$ \alpha_i - 1 $ 
\\
\hline
$ \lambda^a_- $ &
1 &
0 &
$ \alpha_a $ &
$ 1 - \alpha_a $ &
$ - \alpha_a $
\\
\hline
$ \lambda^{\overline{a}}_- $ &
$ -1 $ &
0 &
$ - \alpha_a $ &
$ \alpha_a - 1 $ &
$ \alpha_a $
\\
\hline
\hline
$ D_i $ &
0 &
0 &
$ - \alpha_i $ &
$ \alpha_i $ &
$ \alpha_i $
\\
\hline
$ \overline{D}_{\overline{\imath}} $ &
0 &
0 &
$ \alpha_i $ &
$ - \alpha_i $ &
$ -\alpha_i $
\\
\hline
$ \partial_a $ &
0 &
0 &
$ - \alpha_a $ &
$ \alpha_a $ &
$ \alpha_a $
\\
\hline
$ \overline{\partial}_{\overline{a}} $ &
0 &
0 &
$ \alpha_a $ &
$ - \alpha_a $ &
$ - \alpha_a $
\\
\hline
$ G $ &
0 &
0 &
$ 1 $ &
$ - 1 $ &
$ - 1 $
\\
\hline
$ \overline{G} $ &
0 &
0 &
$ - 1 $ &
$ 1 $ &
$ 1 $
\\
\hline
$ G_a $ &
0 &
0 &
$ 1 - \alpha_a $ &
$ \alpha_a - 1 $ &
$ \alpha_a - 1 $
\\
\hline
$ \overline{G}_{\overline{a}} $ &
0 &
0 &
$ \alpha_a - 1 $ &
$ 1 - \alpha_a $ &
$ 1 - \alpha_a $
\\
\hline
$ F_a = \partial_a G + G_a $ &
0 &
0 &
$ 1 - \alpha_a $ &
$ \alpha_a - 1 $ &
$ \alpha_a - 1 $
\\
\hline
$ 
   \overline{F}_{\overline{a}} = \overline{\partial}_{\overline{a}} \, \overline{G} 
 + \overline{G}_{\overline{a}} 
$ &
0 &
0 &
$ \alpha_a - 1 $ &
$ 1 - \alpha_a $ &
$ 1 - \alpha_a $
\\
\hline  
\end{tabular}
\caption{Charges when $ F_a = \partial_a G + G_a $.}
\label{new-charges}
\end{center}
\end{table}
\noindent
expressed in terms of the parameters
\begin{equation}
\alpha_i 
  = n_i / d 
  = 
  - m_{\overline{\imath}} / d \, ,
\qquad
\alpha_a 
  = n_a / d 
  = 
  - m_{\overline{a}} / d \, ,  
\end{equation}
which yield a 
$ U(1)_L \times U(1)_R $-invariant action.
On the (2,2) locus, we have
\[
\begin{aligned}
Q^{\prime}_L 
\bigl(
       \phi^i 
\bigr)
&= Q^{\prime}_L
      \bigl(
             \psi^{i}_+   
      \bigr) ,
\quad      
Q^{\prime}_L 
\bigl(
       \lambda^i_- 
\bigr)
  = Q^{\prime}_L
      \bigl(
             \phi^{i}   
      \bigr)
  + 1 \, ,
\\  
Q^{\prime}_R 
\bigl(
       \phi^i 
\bigr)
 &= Q^{\prime}_R
      \bigl(
             \lambda^{i}_-  
      \bigr) ,
\quad      
Q^{\prime}_R
\bigl(
       \psi^i_+ 
\bigr)
  = Q^{\prime}_R
      \bigl(
             \phi^{i}   
      \bigr)
  + 1 \, .
\end{aligned}
\]
Off of this locus, although one has a pair of 
$ U(1) $
symmetries, only
$ U(1)_R $
is an R-symmetry.
The twisting is achieved by tensoring the fields with
\[
K^{- Q^{\prime}_R / 2}_{\Sigma} 
\otimes
\overline{K}^{\, Q^{\prime}_L / 2}_{\Sigma} \, .
\]
Recall that for a Riemann surface 
$ \Sigma $
of genus 
$ g $, the degree of the canonical bundle is 
$ 2 g - 2 $.
It follows that, for the bundles 
$ K^{- Q^{\prime}_R / 2}_{\Sigma} $
and
$ \overline{K}^{\, Q^{\prime}_L / 2}_{\Sigma} $
to be well-defined,
$ d $
must
divide
$ g - 1 $,
i.e.
\begin{equation}
\label{genus-constraint}
g = 1
  + k
    d \, ,
\qquad
k = 0,1,2,\ldots \, .    
\end{equation}
This genus issue is well understood; more details can be found in
\cite{GuffinSharpe:A-twisted, Witten:Algebraic, Witten:The-N-matrix}.
Table
\ref{new-charges}
gives a classification of the R-symmetries in the models we are discussing in terms of the charges
$ Q^{\prime}_{L} $
and
$ Q^{\prime}_{R} $.
\subsection{\label{subsection:Examples}Examples}
Let us consider some examples in which
$ \mathcal{E} $
is a deformation of 
$ TX $.
For such examples, the anomaly cancellation conditions
\eqref{anomaly-cancellation-conditions}
are satisfied.

As a first example, consider the case in which
$ X $ 
is a complex affine space
and
$ G $
is a Fermat polynomial: 
\begin{exa}

\label{example-1}
Let
$ X = \mathbf{C}^d $,
and
\[
G = \left(
             \phi^1
      \right)^d
    + \cdots
    + \left(
             \phi^d
      \right)^d \, .
\]
Thus, 
\[
\left(
       n_1,\ldots,n_d
\right) 
  = \left(
           1,\ldots,1
    \right) \, ,
\]
\[
G_a    
  = \left(\phi^a\right)^{d - 1} \, ,
\qquad
a = 1,\ldots,d \, ,
\]
and
\[
\alpha_a 
  = \frac{n_a}{d}
  = \frac{1}{d} \, ,
\qquad
a = 1,\ldots,d \, .    
\]
The twist described in this example can be defined on worldsheets of genus 
$ g $
given by
\eqref{genus-constraint}. 
\end{exa}

As a second example, consider the case in which
$ X $
is a complex projective space and
$ G $
is a Fermat polynomial with zero locus defining a hypersurface in 
$ X $: 
\begin{exa}

\label{example-2}
Let
$ X = \mathbf{CP}^{d-1} $,
and
\[
G = \left(
             \phi^1
      \right)^d
    + \cdots
    + \left(
             \phi^d
      \right)^d \, .
\]
Thus, 
\[
\left\{
         G = 0
\right\}
 \in
    \mathbf{CP}^{d-1}[d] \, ,
\]
\[
\left(
       n_1,\ldots,n_d
\right) 
  = \left(
           1,\ldots,1
    \right) \, ,
\]
\[
G_a    
  = \left(\phi^a\right)^{d - 1} \, ,
\qquad
\phi^a \, G_a = 0 \, ,  
\qquad
a = 1,\ldots,d \, ,
\]
and
\[
\alpha_a 
  = \frac{n_a}{d}
  = \frac{1}{d} \, ,
\qquad
a = 1,\ldots,d \, .    
\]
The twist described in this example can be defined on worldsheets of genus
$ g $
given by
\eqref{genus-constraint}.
\end{exa}

As a final example, consider the case in which
$ X $
is a weighted complex projective space and the zero locus of
$ G $
is a hypersurface in that space: 
\begin{exa}

\label{example-3}
Let
$ X = \mathbf{WCP}^3_{12,8,7,9} $,
and
\[
G = \left(
             \phi^1
      \right)^3
    + \phi^1
      \left(
             \phi^2
      \right)^3
    + \phi^2
      \left(
             \phi^2
      \right)^4
    + \left(
             \phi^4
      \right)^4 \, .
\]
Thus, 
\[
\left\{
         G = 0
\right\}
 \in
    \mathbf{WCP}^3_{n_1,n_2,n_3,n_4}[d]  
  = \mathbf{WCP}^3_{12,8,7,9}[36] \, ,
\]
\[
G_a    
  = \left(\phi^a\right)^{\frac{d}{n_a} - 1} \, ,
\qquad
\phi^a \, G_a = 0 \, ,
\qquad
a = 1,\ldots,4 \, ,
\]
and
\[
\begin{aligned}
\alpha_1 
 &= \frac{n_1}{d}
  = \frac{12}{36}
  = \frac{1}{3} \, ,
\\[1ex]
\alpha_2 
 &= \frac{n_2}{d}
  = \frac{8}{36}
  = \frac{2}{9} \, ,
\\[1ex]
\alpha_3 
 &= \frac{n_3}{d}
  = \frac{7}{36} \, ,
\\[1ex]
\alpha_4 
 &= \frac{n_4}{d}
  = \frac{9}{36}
  = \frac{1}{4} \, .  
\end{aligned}
\]
The twist described in this example can be defined on worldsheets of genus 
$ g $
given by
\eqref{genus-constraint}:
\[
\begin{aligned}
g = 1
  + k
    d
 &= 1
  + k
    \left( 
           36 
    \right) \, ,
\qquad
k = 0,1,2,\ldots
\\
 &= 1,37,73,109,\ldots \, .              
\end{aligned}
\]
\end{exa}
\section{\label{section:Curvature}Curvature constraints}

The action
\eqref{action:Q-exact+non-exact}
is invariant on-shell under the supersymmetry transformations
\eqref{SUSY}
up to a total derivative.
It was claimed in
\cite{GaravusoSharpe:Analogues}
that requiring this invariance
when the superpotential
\eqref{superpotential}
is not holomorphic imposes a constraint which relates the nonholomorphic parameters of the superpotential to the Hermitian curvature. 
This curvature constraint, along with an additional constraint imposed by supersymmetry, was derived in
\cite{Garavuso:Curvature, Garavuso:Nonholomorphic}.
Let us now briefly review the key steps of this derivation; see
\cite{Garavuso:Curvature, Garavuso:Nonholomorphic}
for more details.

Since 
$ \delta f = -i \alpha_- \{ Q, f \} $, 
where 
$ f $ 
is any field, the 
$ Q $-exact part of
\eqref{action:Q-exact+non-exact}
is
$ \delta $-exact 
and hence
$ \delta $-closed.
For the non-exact term of
\eqref{action:Q-exact+non-exact}
involving
$ \Phi^*(K) $,
note that
\[
\int_{\Sigma}
\Phi^*(K)
  = \int_{\Phi(\Sigma)} K
  = \int_{\Phi(\Sigma)}
    \left[
         - i 
           \left(
                  g_{i \overline{\imath}}
                + i
                  B_{i \overline{\imath}} 
           \right)
    \right]       
    d \phi^i
    \wedge
    d \phi^{\overline{\imath}}
\]
and
$ K $
satisfies
\[
\partial K
 =
 - i \, 
   \partial_k
   \left(
          g_{i \overline{\imath}}
        + i
          B_{i \overline{\imath}} 
   \right)
    d \phi^k
    \wedge
    d \phi^i
    \wedge
    d \phi^{\overline{\imath}}
 = 0 \, .    
\]
Thus,
\begin{equation}
\delta
\left[ 
       \Phi^*(K)
\right]       
  = \left[
           \Phi^*(K)
    \right]_k
    \delta
    \phi^k
  = 0 \, .         
\end{equation}
It remains to consider the non-exact expression of
\eqref{action:Q-exact+non-exact}
involving
\[
  \psi_+^{\overline{\imath}}
  \lambda_-^{\overline{a}} 
  \overline{D}_{\overline{\imath}} \overline{F}_{\overline{a}}
- \psi_+^i 
  \lambda_-^{a} 
  D_i F_a \, .
\]
First, we compute
 
\begin{align}
\label{delta-non-exact-2a}
\delta
\left(
       \psi^{\overline{\imath}}_+      
       \lambda^{\overline{a}}_-       
       \overline{D}_{\overline{\imath}}
       \overline{F}_{\overline{a}}
\right)
  &= \left(
          - \alpha_-
            \partial_z \phi^{\overline{\imath}}
     \right)       
     \lambda^{\overline{a}}_- \,
     \overline{D}_{\overline{\imath}} \,
     \overline{F}_{\overline{a}}      
   - \psi_+^{\overline{\imath}}
     \lambda_-^{\overline{a}}
     A^{\overline{b}}_{\overline{\imath} \, \overline{a}, k}
     \left( 
            i
            \alpha_-
            \psi^k_+                   
     \right)                   
     \overline{F}_{\overline{b}} 
\nonumber
\\
  &\phantom{=}
   +\psi^{\overline{\imath}}_+
    \lambda_-^{\overline{a}}
    \left(
           \partial_i
           \overline{D}_{\overline{\imath}} 
           \overline{F}_{\overline{a}}              
         + F_{i \overline{\imath} \, a \overline{a}} \,
           h^{a \overline{b}} \,
           \overline{F}_{\overline{b}}  
    \right)
    \left(
           i
           \alpha_-
           \psi^i_+
    \right).                         
\end{align}
Now, we compute

\begin{align}
\label{delta-non-exact-2b}
\delta
\left(
    - \psi^i_+ 
       \lambda^a_-
       D_i F_a
\right)
  &= 
   - \,
    \alpha_-
    \overline{F}_{\overline{a}} \,
    D_z \lambda^{\overline{a}}_-
  + \left(
            i
            \alpha_-
            h^{a \overline{b}} \,
            \overline{F}_{\overline{b}}
     \right)    
     F_{i \overline{\imath} a \overline{a}} \,
     \psi^i_+
     \psi^{\overline{\imath}}_+
     \lambda^{\overline{a}}_- 
\nonumber
\\[1ex] 
  &=
     \left(
           \alpha_-
           \partial_z \phi^{\overline{\imath}}
     \right)       
     \lambda^{\overline{a}}_- \,
     \overline{D}_{\overline{\imath}} \,
     \overline{F}_{\overline{a}}
   - \alpha_-
     \partial_z \!
     \left(
            \overline{F}_{\overline{a}} \,
            \lambda^{\overline{a}}_-
     \right)     
   + \alpha_- 
     \overline{F}_{\overline{a}, k} \,
     \partial_z \phi^k
     \lambda^{\overline{a}}_-
\nonumber
\\     
 &\phantom{=}    
   + \psi^{\overline{\imath}}_+
     \lambda^{\overline{a}}_- \,
     A^{\overline{b}}_{\overline{\imath} \, \overline{a}, k} \,
     \left(
            i
            \alpha_-
            \psi^k_+
     \right)
     \overline{F}_{\overline{b}} \, ,                                                    
\end{align}
where we have used the 
$ \lambda^a_- $
equation of motion
\eqref{eom:lambda-super-a-sub-minus}
in the first step
and
$
  F_{i \overline{\imath} a \overline{a}}
= h_{a \overline{b}} \,
  A^{\overline{b}}_{\overline{\imath} \, \overline{a}, i}  
$
in the last step.
It follows that
\eqref{delta-non-exact-2b}
cancels
\eqref{delta-non-exact-2a}
up to a total derivative, i.e.
\begin{equation}
\delta
\left(
    - \psi^i_+ 
       \lambda^a_-
       D_i F_a
\right)
  = 
  - \,
    \delta
    \left(
           \psi^{\overline{\imath}}_+
           \lambda^{\overline{a}}_- \,
           \overline{D}_{\overline{\imath}}
           \overline{F}_{\overline{a}}
    \right)
  - \alpha_-
    \partial_z \!
    \left(
           \overline{F}_{\overline{a}} \,
            \lambda^{\overline{a}}_-
    \right),  
\end{equation}
when both the curvature constraint
\begin{equation}
\label{constraint-curvature}
  \partial_i
  \overline{D}_{\overline{\imath}} 
  \overline{F}_{\overline{a}}              
+ F_{i \overline{\imath} \, a \overline{a}} \,
  h^{a \overline{b}} \,
  \overline{F}_{\overline{b}}
= 0             
\end{equation}
and the constraint
\begin{equation}
\label{constraint-supplementary}
\overline{F}_{\overline{a}, k} \,
\partial_z \phi^k
\lambda^{\overline{a}}_-
  = 0    
\end{equation}
are satisfied.

Curvature constraints have been used in
\cite{GaravusoSharpe:Analogues}
to establish properties of analogues of pullbacks of Mathai-Quillen forms.
These analogues arise in the correlation functions of the corresponding A-twisted or B-twisted heterotic Landau-Ginzburg models.
The analogue most relevant to this paper, i.e. a deformation of the pullback of a Mathai-Quillen form, is discussed in appendix
\ref{section:Deformation-pullback-Mathai-Quillen}.
\section{\label{section:Physical}Physical realization of deformation of the pullback of a Mathai-Quillen form}

Let us now describe how the deformation
$ \omega_{\delta s}(\mathcal{G},\nabla) $,
given by
\eqref{omega-delta-s},
of the pullback
$ s^* u(\mathcal{G}, \nabla) $,
given by
\eqref{MQ-Euler},
of a Mathai-Quillen form
$ u(\mathcal{G}, \nabla) $,
given by
\eqref{MQ-form},
may arise in the class of A-twisted heterotic Landau-Ginzburg models discussed in this paper.  

Mathematically, the tangent bundle to 
$ Y \equiv \{ s = 0 \} $,
$ s = (s_p)$, 
is defined by the kernel in the short exact sequence
\begin{displaymath}
0 \: \longrightarrow \: 
T Y \: \longrightarrow \: TM |_Y \: \stackrel{(D_i s_p)}{
\longrightarrow} \:
{\cal G}|_Y \: \longrightarrow \: 0 .
\end{displaymath}
A deformation of the tangent bundle above is defined by
\begin{displaymath}
0 \: \longrightarrow \: 
{\mathcal E'} \: \longrightarrow \:
 TM |_Y \: \stackrel{(D_i s_p + (\delta s)_{i p})}{
\longrightarrow} \:
{\cal G}|_Y \: \longrightarrow \: 0 ,
\end{displaymath}
where the
$(\delta s)_{i p}$
define the deformation.

The action of the A-twisted heterotic Landau-Ginzburg model that RG flows to a nonlinear sigma model with tangent bundle deformation above is given by
\cite{GuffinSharpe:A-twistedheterotic}

\begin{align}
S &= 2 t \int_{\Sigma} d^2 z
     \left[
            \frac{1}{2}
            \left(
                   g_{\mu\nu} 
                 + i B_{\mu\nu}
            \right)
            \partial_z \phi^{\mu}
            \overline{\partial}_{\overline{z}} \phi^{\nu}
          + i g_{\overline{a}a} \psi^{\overline{a}}_+ \overline{D}_{\overline{z}} \psi^a_+
          + i g_{b\overline{b}} \lambda^b_- D_z \lambda^{\overline{b}}_-
     \right.
\nonumber
\\
  &\phantom{= 2 t \int_{\Sigma} d^2 z \left[ \right.}
     \biggl.
          + R_{a\overline{a}b\overline{b}}
            \psi^a_+
            \psi^{\overline{a}}_+
            \lambda^b_-
            \lambda^{\overline{b}}_- 
          + g^{a\overline{a}}
            F_a 
            \overline{F}_{\overline{a}}
          + \psi^a_+ 
            \lambda^b_- 
            D_a F_b
          + \psi^{\overline{a}}_+ 
            \lambda^{\overline{b}}_- 
            \overline{D}_{\overline{a}} \overline{F}_{\overline{b}}        
     \biggr],
\end{align}
with target space
\begin{displaymath}
X \: = \: {\rm Tot} \left( {\cal G}^* \: \stackrel{\pi}{\longrightarrow} \:
M \right)
\end{displaymath}
and gauge bundle
$ {\cal E} = TX $,
where
\[
F_a 
   = (F_p, F_i)
   = \left(
            s_p, \phi^p (D_i s_p + (\delta s)_{i p})
     \right) \, ,
\qquad       
\overline{F}_{\overline{a}} 
   = \left(
            \overline{F}_{\overline{p}}, \overline{F}_{\overline{\imath}}
     \right)
   = \left(
            \overline{s}_{\overline{p}}, 
            \phi^{\overline{p}} 
            \left( 
                   \overline{D}_{\overline{\imath}} \overline{s}_{\overline{p}} 
                 + (\delta \overline{s})_{\overline{\imath} \, \overline{p}}
            \right)
     \right) \, ,
\]
\[
\begin{aligned}
D_a F_b 
  &= \partial_a F_b 
   - \Gamma^c_{ab} 
     F_c \, ,
\quad
&\overline{D}_{\overline{a}}  \overline{F}_{\overline{b}}
  &= \overline{\partial}_{\overline{a}} \overline{F}_{\overline{b}}
   - \Gamma^{\overline{c}}_{\overline{a}\overline{b}}
     \overline{F}_{\overline{c}} \, ,    
\\
\overline{D}_{\overline{z}} \psi^a_+ 
   &= \overline{\partial}_{\overline{z}} \psi^a_+ 
    + \overline{\partial}_{\overline{z}} \phi^b \Gamma^a_{bc} \psi^c_+ \, ,
\quad
&D_z \lambda^{\overline{b}}_-
  &= \partial_z \lambda^{\overline{b}}_- 
   + \partial_z \phi^{\overline{a}} 
     \Gamma^{\overline{b}}_{\overline{a} \, \overline{c}} \lambda^{\overline{c}}_- \, ,          
\end{aligned}
\]
and
\[
\begin{aligned}
\psi^i_+ 
  &\equiv \chi^i \!
  &\in \
  &\Gamma
   \left(  
          \Phi^* \!
          \left(
                 T^{1,0} M
          \right)       
  \right),
\quad          
&\lambda^i_-
  &\equiv \lambda^i_{\overline{z}} \!
  &\in \
  &\Gamma
   \left( 
          \overline{K}_{\Sigma}
          \otimes
          \left( 
                 \Phi^* \!
                 \left(
                        T^{0,1} M
                 \right)       
         \right)\!^{\vee}
  \right),       
\\
\psi^{\overline{\imath}}_+
  &\equiv \psi^{\overline{\imath}}_z \!
  &\in \
  &\Gamma
   \left(
          K_{\Sigma}
          \otimes
          \left( 
                 \Phi^* \!
                 \left(
                        T^{1,0} M
                 \right)       
          \right)\!^{\vee}
  \right),
\quad     
&\lambda^{\overline{\imath}}_-
  &\equiv \lambda^{\overline{\imath}} \!
  &\in \
  &\Gamma
   \left( 
          \Phi^* \!
          \left(
                 T^{0,1} M
          \right)       
   \right),
\\
\psi^p_+
  &\equiv \psi^p_z \!
  &\in \ 
  &\Gamma
   \left(
          K_{\Sigma}
          \otimes
          \Phi^* \,
          T^{1,0}_{\pi}
   \right),
\quad
&\lambda^p_-
  &\equiv \lambda^p \!
  &\in \ 
  &\Gamma
   \left(
          \left(
                 \Phi^* \,
                 T^{0,1}_{\pi}
          \right)\!^{\vee}         
   \right),
\\
\psi^{\overline{p}}_+
  &\equiv \chi^{\overline{p}} \!
  &\in \ 
  &\Gamma
   \left(
          \left(
                 \Phi^* \,
                 T^{1,0}_{\pi}
          \right)\!^{\vee} 
   \right),
\quad   
&\lambda^{\overline{p}}_-
  &\equiv \lambda^{\overline{p}}_{\overline{z}} \!
  &\in \ 
  &\Gamma
   \left(
          \overline{K}_{\Sigma}
          \otimes
          \Phi^* \,
          T^{0,1}_{\pi}               
   \right),
\\
\phi^p 
  &\equiv p_z \!
  &\in \
  &\Gamma
   \left(
          K_{\Sigma} 
          \otimes 
          \Phi^* \,
          T^{1,0}_{\pi}
   \right),
\quad
&\phi^{\overline{p}} 
  &\equiv \overline{p}_{\overline{z}} \! 
  &\in \
  &\Gamma
   \left(
          \overline{K}_{\Sigma} 
          \otimes 
          \Phi^* \,
          T^{0,1}_{\pi}
   \right).                                              
\end{aligned}
\]

If we restrict to zero modes on a genus zero worldsheet, in the degree zero sector
we find the following interactions among zero modes:

\begin{align} 
     g^{\overline{p}p} &
     \overline{F}_{\overline{p}}
     F_p
   + \chi^i
     \lambda^p
     D_i F_p
   + \chi^{\overline{p}}
     \lambda^{\overline{\imath}} \,
     \overline{D}_{\overline{p}}
     \overline{F}_{\overline{\imath}}
   + R_{i\overline{p}p\overline{\imath}}
     \chi^i
     \chi^{\overline{p}}
     \lambda^p
     \lambda^{\overline{\imath}}      
\nonumber
\\[1ex]     
  &= g^{\overline{p}p}
     \overline{s}_{\overline{p}}
    s_p 
   + \chi^i
     \lambda^p
     D_i s_p           
   + \chi^{\overline{p}}
     \lambda^{\overline{\imath}}
     \left( 
            \overline{D}_{\overline{\imath}} \overline{s}_{\overline{p}} 
          + (\delta \overline{s})_{\overline{\imath} \overline{p}}
     \right)       
   + R_{i\overline{p}p\overline{\imath}}
     \chi^i
     \chi^{\overline{p}} 
     \lambda^p
     \lambda^{\overline{\imath}}\, .
\nonumber
\end{align}
If we now complex conjugate so as to relate the heterotic expression above to
standard mathematics conventions, we find

\begin{align}
     g^{p\overline{p}} &
     s_p
     \overline{s}_{\overline{p}}
   + \chi^{\overline{\imath}}
     \lambda^{\overline{p}} \,
     \overline{D}_{\overline{\imath}} \overline{s}_{\overline{p}}                        
   + \chi^{p}
     \lambda^i
     \left( 
            D_{i} s_p 
          + (\delta s)_{i p}
     \right)       
   + R_{\overline{\imath}p\overline{p}i}
     \chi^{\overline{\imath}}
     \chi^{p} 
     \lambda^{\overline{p}}
     \lambda^{i}
\nonumber
\\[1ex]
  &= g^{p\overline{p}}
     s_p
     \overline{s}_{\overline{p}}
   + \rho^p D s_p
   + \overline{D} \overline{s}_{\overline{p}} \, \rho^{\overline{p}}        
   + \rho^{\overline{p}}
     \mathcal{R}_{\overline{p}p}
     \rho^p
   + \rho^{p}
     d \phi^{i} \, 
     (\delta s)_{i p}
\nonumber
\\[1ex]
  &= \left(
            s_p e^p,
            \overline{s}_{\overline{p}} e^{\overline{p}}
     \right)_{\cal G}
   + \left\langle
            \rho^{p'} f_{p'},
            D s_p e^p
     \right\rangle_{\cal G}
   + \left\langle
            \overline{D} \overline{s}_{\overline{p}} e^{\overline{p}},
            \rho^{\overline{p}'} f_{\overline{p}'}
     \right\rangle_{\cal G}
\nonumber
\\
  &\phantom{=}
   + \left(
            \rho^{\overline{p}'} f_{\overline{p}'},
            f^{\overline{p}}
            \left(
                   \mathcal{R}_{\overline{p}p} f^p, 
                   \rho^p f_p
            \right)_{{\cal G}^{\vee}}
     \right)_{{\cal G}^{\vee}}
   + \left\langle
            \rho^{p'} f_{p'},
            d \phi^{i} \,
            (\delta s)_{i p}
            e^{p} 
     \right\rangle_{\cal G}
\nonumber
\\[1ex]
  &= {\cal A} 
   + \left\langle
            \rho^{p'} f_{p'},
            d \phi^{i} \,
            (\delta s)_{i p}
            e^{p} 
     \right\rangle_{\cal G}
\nonumber
\\[1ex]
  &= {\cal A}_{\delta s} \, ,                                         
\end{align}
which is minus the exponent of
\eqref{omega-delta-s}.
The deformation
$ \omega_{\delta s}(\mathcal{G},\nabla) $
will appear in the corresponding correlation functions.
Explicitly, from the discussion in appendix
\ref{section:Deformation-pullback-Mathai-Quillen}, we see that
\begin{equation}
\langle
        \widetilde{\mathcal{O}}_1
        \cdots
        \widetilde{\mathcal{O}}_k
\rangle
  \propto  
    \int_X
    \widetilde{\mathcal{O}}_1
    \wedge
    \cdots
    \wedge
    \widetilde{\mathcal{O}}_k
    \wedge
    \omega_{\delta s}(\mathcal{G}, \nabla)
  = \int_Y
    \mathcal{O}_1
    \wedge
    \cdots
    \wedge
    \mathcal{O}_k \, ,
\end{equation}
where
\[ 
  \widetilde{\mathcal{O}}_1
  \wedge
  \cdots
  \wedge
  \widetilde{\mathcal{O}}_k
    \in
    H^{\dim{M} - \textrm{rk} \, \mathcal{G}}
    \left(
           M,
           \wedge^{\textrm{rk} \, TM -  \textrm{rk} \, \mathcal{G}}
           \mathcal{G}^{\vee}
    \right)
\]
and
$
  \widetilde{\mathcal{O}}
  \in
  H^{\bullet}
  \left(
         M,
         \wedge^{\bullet}
         (TM)^{\vee}
  \right) 
$
is a lift of
$
  \mathcal{O}
  \in
  H^{\bullet}
  \left(
         Y,
         \wedge^{\bullet}
         \mathcal{\mathcal{E}}^{\prime \vee}
  \right)
$.
\section{\label{section:Summary}Summary and outlook}

We have studied certain aspects of A-twisted heterotic Landau Ginzburg models on a 
K\"{a}hler
variety
$ X $
with gauge bundle
$ \mathcal{E} $,
superpotential
\eqref{superpotential}
\[
W = \Lambda^a
    F_a \, ,
\]
and
$ E^a \equiv 0 $.
Table
\ref{new-charges}
provides a classification of the R-symmetries which allow the A-twist to be implemented when
$ \mathcal{E} $
is either a deformation of 
$ TX $
or a deformation of a sub-bundle of
$ TX $.
Some anomaly-free examples were provided in section
\ref{subsection:Examples}.
When the superpotential is not holomorphic, supersymmetry imposes the curvature constraint
\eqref{constraint-curvature}
\[
  \partial_i
  \overline{D}_{\overline{\imath}} 
  \overline{F}_{\overline{a}}              
+ F_{i \overline{\imath} \, a \overline{a}} \,
  h^{a \overline{b}} \,
  \overline{F}_{\overline{b}}
= 0             
\]
and the constraint
\eqref{constraint-supplementary}
\[
\overline{F}_{\overline{a}, k} \,
\partial_z \phi^k
\lambda^{\overline{a}}_-
  = 0 \, .     
\]
The curvature constraint
\eqref{constraint-curvature}
was used in
\cite{GaravusoSharpe:Analogues}
to establish properties of the deformation
$ \omega_{\delta s}(\mathcal{G},\nabla) $,
given by
\eqref{omega-delta-s},
of the pullback
$ s^* u(\mathcal{G}, \nabla) $,
given by
\eqref{MQ-Euler},
of a Mathai-Quillen form
$ u(\mathcal{G}, \nabla) $,
given by
\eqref{MQ-form}.
In section
\ref{section:Physical},
we described how
$ \omega_{\delta s}(\mathcal{G},\nabla) $
may arise in the class of heterotic Landau-Ginzburg models studied in this paper. 

It would be interesting to consider A-twisted and B-twisted heterotic Landau-Ginzburg models with more general Grassmann-odd superpotentials.
For example, one may consider the superpotential
\cite{GuffinSharpe:A-twisted}
\[
W = \Lambda^a
    \Lambda^b
    \Lambda^c
    F_{abc} \, .
\]
If this more general superpotential is not holomorphic, then supersymmetry should impose a curvature constraint analogous to
\eqref{constraint-curvature}
and a constraint analogous to
\eqref{constraint-supplementary}.
These new constraints could be derived using arguments similar to those used in
\cite{Garavuso:Curvature, Garavuso:Nonholomorphic}.
Furthermore, using arguments similar to those used in
\cite{GaravusoSharpe:Analogues},
the new curvature constraint could then be used to establish properties of new 
analogues of pullbacks of Mathai-Quillen forms which arise in the correlation functions of the corresponding A-twisted or B-twisted heterotic Landau-Ginzburg models.
We leave a detailed study of this to future work.
\acknowledgments{The author thanks Mauricio Romo for useful discussions.}
\begin{appendix}

\section{\label{section:Review-MQ-formalism}Review of Mathai-Quillen formalism}

Consider an oriented vector bundle 
$ \mathcal{G} \overset{\pi}{\longrightarrow} M $ 
of real rank
$ r = 2m $,
with standard fiber 
$ V $,
where 
$ M $
is an oriented closed manifold of real dimension 
$ n \geq r $.
Suppose that
$ \mathcal{G} $
has Euclidean metric
$ (\cdot,\cdot)_{\mathcal{G}} $
and compatible connection
$ \nabla $.
Under these circumstances, the
\emph{Mathai-Quillen formalism}
\cite{MathaiQuillen:Superconnections,
BerlinGetzlerVergne:Heat,
Kalkman:BRST,
Blau:The-Mathai-Quillen,
Wu:On-the-Mathai-Quillen,
CordesMooreRamgoolam:Lectures,
Wu:Mathai-Quillen} 
provides an explicit representative
$ u(\mathcal{G},\nabla) $
of the Thom class of
$ \mathcal{G} $.
Furthermore, the pullback 
$ s^* u(\mathcal{G},\nabla) $
of $ u(\mathcal{G},\nabla) $
by any section 
$ s: M \rightarrow \mathcal{G} $
of
$ \mathcal{G} $
is a representative of the Euler class of 
$ \mathcal{G} $.
Let us review the formalism in more detail.
\subsection{Conventions}

Our conventions for
$ M $, 
$ \mathcal{G} $, and the dual 
$ \mathcal{G}^{\vee} $
of
$ \mathcal{G} $
are as follows.
The exterior derivatives on
$ M $
and
$ \mathcal{G} $
are respectively denoted by
$ d $
and
$ d^{\mathcal{G}} $.
We choose local coordinates
$ \phi^I $
on
$ M $,
where
$ I = 1,\ldots,n $.
The connection on
$ \mathcal{G} $
is then given by
$ \nabla = d \phi^I \nabla_I $.
In terms of this connection, the curvature 2-form on
$ \mathcal{G} $
is given by $ \mathcal{R} = \nabla^2 $.
We choose a local oriented orthonormal frame
$ \{ e_A \} $
for
$ \mathcal{G} $
and let
$ \{ f^A \} $
be the dual coframe, where
$ A = 1,\ldots,r $.
The section
$ s $
may thus be expressed as $ s = s^A e_A $.
Similarly, we write
$ \rho = \rho_A f^A $,
where the
$ \rho_A $
are anticommuting orthonormal coordinates on 
$ \mathcal{G}^{\vee} $.
The dual pairing on
$ \mathcal{G} $
is denoted by
$ \langle \cdot, \cdot \rangle_{\mathcal{G}} $.
Finally, the metric on
$ \mathcal{G}^{\vee} $
is denoted by
$ (\cdot,\cdot)_{\mathcal{G}^{\vee}} $. 

Now, consider the pullback bundle 
$ \pi^* \mathcal{G} \rightarrow \mathcal{G} $,
i.e. the bundle over 
$ \mathcal{G} $
whose fiber at 
$ g \in \mathcal{G} $ is 
$ (\pi^* \mathcal{G})_g = \mathcal{G}_{\pi(g)} $.
This bundle has Euclidean metric 
$ \pi^* (\cdot,\cdot)_{\mathcal{G}} \equiv (\cdot,\cdot)_{\pi^* \mathcal{G}} $,
compatible connection 
$ \pi^* \nabla \equiv \widetilde{\nabla} $,
curvature 2-form 
$ \pi^* \mathcal{R} \equiv \widetilde{\mathcal{R}} $,
local oriented orthonormal frame 
$ \{ \pi^* e_A \} \equiv \{ \tilde{e}_A \} $, and 
\emph{tautological section}
$ \tilde{x} = \tilde{x}^A \tilde{e}_A $.
(The tautological section of $ \pi^* \mathcal{G} \rightarrow \mathcal{G} $ is the section which maps a point $ g \in \mathcal{G} $ to $ (g,g) \in \pi^* \mathcal{G} $.)
The dual bundle 
$ (\pi^* \mathcal{G})^{\vee} \rightarrow \mathcal{G} $ 
has coframe 
$ \{ (\pi^{\vee})^* f^A \} \equiv \{ \tilde{f}^A \} $ 
and metric 
$ (\cdot,\cdot)_{(\pi^* \mathcal{G})^{\vee}} $.
We write 
$ \tilde{\rho} = \tilde{\rho}_A \tilde{f}^A $, 
where the
$ \tilde{\rho}_A \equiv (\pi^{\vee})^* \rho_A $
are anticommuting orthonormal coordinates on
$ (\pi^* \mathcal{G})^{\vee} $.
The dual pairing on
$ \pi^* E $
is denoted by
$ \langle \cdot, \cdot \rangle_{\pi^* \mathcal{G}} $.

\subsection{Mathai-Quillen Thom class representative}

Consider the \emph{Mathai-Quillen form}
\begin{equation}
\label{MQ-form}
u(\mathcal{G},\nabla)
   = a_r \int d \tilde{\rho} \,
     \exp
     \left(
          - \tilde{\mathcal{A}}
     \right), 
\end{equation}
where
\begin{equation}
\label{a-sub-r}
a_r 
   = \frac{  (-1)^{ \frac{r(r+1)}{2} }  }
          {  (2 \pi)^{ \frac{r}{2} }  }
\end{equation}
and
\begin{equation}
\tilde{\mathcal{A}} 
   = \frac{1}{2} 
     \Bigl( 
            \tilde{x}, \tilde{x} 
     \Bigr)_{\pi^* \mathcal{G}} 
   + \left\langle 
            \widetilde{\nabla} \tilde{x}, \tilde{\rho} 
     \right\rangle_{\pi^* \mathcal{G}}
   + \frac{1}{2} 
     \left(
            \tilde{\rho}, \widetilde{\mathcal{R}} \tilde{\rho} 
     \right)_{(\pi^* \mathcal{G})^{\vee}}.                
\end{equation}
We wish to show that this form satisfies the following definition.
\begin{defi}
\label{u(mathcal-G)}
A representative of the Thom class of 
$ \mathcal{G} $
is a
$ d^{\mathcal{G}} $-closed
differential form
$ u(\mathcal{G}) \in \Omega^r(\mathcal{G}) $
such that
$ \int_V u(\mathcal{G}) = 1 $.
\end{defi}
\begin{prop}
\label{prop-MQ}
\renewcommand{\theenumi}{\theproposition\arabic{enumi}}
The Mathai-Quillen form
$ u(\mathcal{G},\nabla) $
satisfies
\begin{enumerate}
\renewcommand*{\labelenumi}{\theenumi}
\renewcommand{\theenumi}{(\roman{enumi})}

\item
\label{prop-MQ-i}
$ u(\mathcal{G},\nabla) \in \Omega^r(\mathcal{G}) \, , $

\item
\label{prop-MQ-ii}
$ d^{\mathcal{G}} u(\mathcal{G},\nabla) = 0 \, , $

\item
$ \int_V u(\mathcal{G},\nabla) = 1 $

\end{enumerate}
and hence is a representative of the Thom class of
$ \mathcal{G} $.
\end{prop}
\begin{proof}
\mbox{}
\begin{enumerate}
\renewcommand*{\labelenumi}{\theenumi}
\renewcommand{\theenumi}{(\roman{enumi})}

\item  
Since
\[
\tilde{\mathcal{A}} 
   \in \overset{2}{\underset{i=0}{\oplus}} \Omega^i
       \left(
              \mathcal{G}, 
              \Lambda^i 
              \left(
                     \pi^* \mathcal{G} 
              \right)^{\vee} 
       \right),                  
\]
it follows that
\[
\exp
\left(
     - \tilde{\mathcal{A}}
\right)
   \in \overset{r}{\underset{i=0}{\oplus}} \Omega^i
       \left(
              \mathcal{G}, 
              \Lambda^i 
              \left(
                     \pi^* \mathcal{G} 
              \right)^{\vee} 
       \right).
\]
However, only the component of
$ e^{- \tilde{\mathcal{A}}} $
in 
$ \Omega^r \left( \mathcal{G}, \Lambda^r \left( \pi^* \mathcal{G}  \right)^{\vee} \right) $
contributes to
$ u(\mathcal{G},\nabla) $. 
Thus,
$ u(\mathcal{G},\nabla) \in \Omega^r(\mathcal{G}) $. 

\item
Since
$ \widetilde{\nabla} $
is compatible with the metric
$ \left( \cdot, \cdot \right)_{\pi^* \mathcal{G}} $,
it follows that
\[
d^{\mathcal{G}} \int d \tilde{\rho} \, 
\tilde{\alpha}
   = \int d \tilde{\rho} \,
     \widetilde{\nabla} 
     \tilde{\alpha} \, ,
\]
where
$
\tilde{\alpha}
   \in \Omega
       \left(
              \mathcal{G}, 
              \Lambda
              \left( 
                     \pi^* \mathcal{G}
              \right)^{\vee}
       \right)
$.
Furthermore,

\begin{align}
\left( 
       \widetilde{\nabla}
     + \tilde{x}_A
       \frac{\partial}{\partial \tilde{\rho}_A}   
\right)
\tilde{\mathcal{A}}
  &= \left(
            \widetilde{\nabla} \tilde{x}, \tilde{x}
     \right)_{\pi^* \mathcal{G}}  
   + \left\langle
            \widetilde{\mathcal{R}} \tilde{x}, \tilde{\rho}
     \right\rangle_{\pi^* \mathcal{G}}
   - \frac{1}{2}
     \left(
            \tilde{\rho}, \widetilde{\nabla} \widetilde{\mathcal{R}} \tilde{\rho} 
     \right)_{\left(\pi^* \mathcal{G}\right)^{\vee}}
\nonumber
\\[1ex]
  &\phantom{=}
   - \left(
            \widetilde{\nabla} \tilde{x}, \tilde{x}
     \right)_{\pi^* \mathcal{G}}  
   - \left\langle
            \widetilde{\mathcal{R}} \tilde{x}, \tilde{\rho}
     \right\rangle_{\pi^* \mathcal{G}}
\nonumber
\\[2ex]
  &= 0 \, ,                            
\end{align}
where we have used the Bianchi identity 
$ \widetilde{\nabla} \, \widetilde{\mathcal{R}} = 0 $.
From these results, we obtain
\[
\begin{aligned}
d^{\mathcal{G}} u(\mathcal{G},\nabla)
  &= a_r \, 
     d^{\mathcal{G}} \int d \tilde{\rho} \, 
     \exp 
     \left( 
          - \tilde{\mathcal{A}} 
     \right)
\\[1ex]
  &= a_r \int d \tilde{\rho} \,
     \widetilde{\nabla} 
     \exp
     \left(
          - \tilde{\mathcal{A}}
     \right)
\\[1ex]
  &= a_r \int d \tilde{\rho} \,
     \left(
            \widetilde{\nabla}
          + \tilde{x}_A
            \frac{\partial}{\partial \tilde{\rho}_A}  
     \right)
     \exp
     \left(
          - \tilde{\mathcal{A}}
     \right)
\\[1ex]
  &= a_r \int d \tilde{\rho} \,
     \left[
          - \left(  
                   \widetilde{\nabla}
                 + \tilde{x}_A
                   \frac{\partial}{\partial \tilde{\rho}_A}
            \right)
            \tilde{\mathcal{A}}         
     \right]
     \exp
     \left(
          - \tilde{\mathcal{A}}
     \right)
\\[1ex]
   &= 0 \, .                      
\end{aligned}
\]
Here, the third equality holds because                  
$ 
\tilde{x}_A 
\left( 
       \partial / \partial \tilde{\rho}_A 
\right)
e^{
  - \tilde{\mathcal{A}}
  }
$
contributes nothing to the Grassmann integral.

\item
\[
\begin{aligned}
\int_V u(\mathcal{G},\nabla)
  &= a_r \int_V
     \exp
     \left[
          - \frac{1}{2} 
            \left( 
                   \tilde{x}, \tilde{x}
            \right)_{\pi^* \mathcal{G}} 
     \right]   
     \int d \tilde{\rho} \,
     \frac{ 
            \left(
                 - d \tilde{x}^A \tilde{\rho}_A 
            \right)^r
           }{r!}   
\\[1ex]       
  &= \frac{1}{  (2 \pi)^{ \frac{r}{2} }  }
     \int_V d \tilde{x}^1 \wedge \cdots \wedge d \tilde{x}^r \, 
     \exp
     \left[
          - \frac{1}{2} 
            \left( 
                   \tilde{x}, \tilde{x}
            \right)_{\pi^* \mathcal{G}} 
     \right]     
\\[1ex]          
  &= 1 \, .       
\end{aligned}
\]

\end{enumerate}
\noindent
Thus, by definition
\ref{u(mathcal-G)}, 
$ u(\mathcal{G},\nabla) $
is a representative of the Thom class of
$ \mathcal{G} $.
\qedhere

\end{proof}
\subsection{Mathai-Quillen Euler class representative}

Now, consider the pullback of the Mathai-Quillen form
$ u(\mathcal{G},\nabla) $
by any section
$ s $
of 
$ \mathcal{G} $.
We write this as
\begin{equation}
\label{MQ-Euler}
s^* u(\mathcal{G},\nabla)
   = a_r
     \int d \rho \,
     \exp 
     \left(
          - \mathcal{A}     
     \right), 
\end{equation}
where
$ a_r $
is given by 
\eqref{a-sub-r}
and
\begin{equation}
\label{mathcal-A}
\mathcal{A} 
  = \frac{1}{2} 
    \left(
           s, s
    \right)_{\mathcal{G}} 
  + \left\langle 
            \nabla s, \rho 
    \right\rangle_{\mathcal{G}}
  + \frac{1}{2} 
    \left(
           \rho, \mathcal{R} \rho
    \right)_{\mathcal{G}^{\vee}}.
\end{equation}
\begin{prop}
The form
$ s^* u(\mathcal{G},\nabla) $
satisfies
\begin{enumerate}

\item[(i)]
$ s^* u(\mathcal{G},\nabla) \in \Omega^r(M) \, , $ 

\item[(ii)]
$ d s^* u(\mathcal{G},\nabla) = 0 \, . $

\end{enumerate}

\end{prop}
\begin{proof}
\mbox{}
\begin{enumerate}

\item[(i)]
The proof is similar to that of proposition
\ref{prop-MQ}\ref{prop-MQ-i}
and uses the fact that
\[
\mathcal{A}
  \in \overset{2}{\underset{i=0}{\oplus}} \Omega^i
      \left(
             \mathcal{G}, 
             \Lambda^i \mathcal{G}^{\vee} 
      \right). 
\]

\item[(ii)]
The proof is similar to that of proposition
\ref{prop-MQ}\ref{prop-MQ-ii}
and uses the results
\[
d \int d \rho \, 
\alpha
   = \int d \rho \,
     \nabla 
     \alpha \, ,           
\]
where
$
\alpha
   \in \Omega
       \left(
              \mathcal{G}, 
              \Lambda \mathcal{G}^{\vee}
       \right)
$, and 
\begin{equation}
\label{nabla-mathcal-A-equation}
\left(
       \nabla 
     + s_A \frac{\partial}{\partial \rho_A} 
\right) \mathcal{A}
   = 0 \, . 
\end{equation}

\end{enumerate}
\end{proof}
\begin{prop}
\label{MQ-independent-s}
The
$ d $-cohomology
class of
$ s^* u(\mathcal{G},\nabla) $
is independent of the section
$ s $.  
\end{prop}
\begin{proof}
Let
$ s_{\tau} = s + \tau s^{\prime} $
be an affine one-parameter family of sections of
$ \mathcal{G} $
and let
\[
\mathcal{A}_{\tau}
   =
     \frac{1}{2} 
     \left(
            s_{\tau}, s_{\tau}
     \right)_{\mathcal{G}} 
   + \left\langle
            \nabla s_{\tau}, \rho
     \right\rangle_{\mathcal{G}}
   + \frac{1}{2} 
     \left(
            \rho, \mathcal{R} \rho
     \right)_{\mathcal{G}^{\vee}}.
\]
Then
\[
\begin{aligned}
\frac{d}{d \tau} 
s^*_{\tau} u(\mathcal{G},\nabla)
  &= a_r \, 
     \frac{d}{d \tau}
     \int d \rho \,     
     \exp 
     \left(
          - \mathcal{A}_{\tau}
     \right)
\\[1ex]          
  &=
   - a_r
     \int d \rho \,
     \Bigl[
            \left(
                   s^{\prime}, s_{\tau}
            \right)_{\mathcal{G}}  
          + \left\langle
                   \nabla s^{\prime}, \rho
            \right\rangle_{\mathcal{G}}   
     \Bigr]          
     \exp 
     \left(
          - \mathcal{A}_{\tau}
     \right)              
\nonumber
\\[1ex]
  &=
   - a_r \,  
     \int d \rho \,
     \left\{
            \left[
                   \nabla 
                 + \left(
                          s_{\tau}
                   \right)_A       
                   \frac{\partial}{\partial \rho_A}
            \right]       
            \left\langle
                   s^{\prime}, \rho
            \right\rangle_{\mathcal{G}}
     \right\}
     \exp 
     \left(
          - \mathcal{A}_{\tau} 
     \right)
\nonumber
\\[1ex]
  &=
   - a_r \,  
     \int d \rho \,
     \left[
            \nabla 
          + \left(
                   s_{\tau}
            \right)_A       
            \frac{\partial}{\partial \rho_A}
     \right]
     \Bigl[       
            \left\langle
                   s^{\prime}, \rho
            \right\rangle_{\mathcal{G}} \,
            \exp 
            \left(
                 - \mathcal{A}_{\tau} 
            \right)
     \Bigr]       
\nonumber
\\[1ex]
  &=
   - a_r \, 
     d
     \int d \rho \,   
     \left\langle
            s^{\prime}, \rho
     \right\rangle_{\mathcal{G}} \,          
     \exp 
     \left(
          - \mathcal{A}_{\tau} 
     \right).                         
\end{aligned}
\]
It follows that
\[
s^*_{\tau_2} u(\mathcal{G},\nabla) - s^*_{\tau_1} u(\mathcal{G},\nabla)
   =
   - a_r \,  
     d \int^{\tau_2}_{\tau_1} d \tau
     \int d \rho \,   
     \left\langle
            s^{\prime}, \rho
     \right\rangle_{\mathcal{G}} \,                 
     \exp 
     \left(
          - \mathcal{A}_{\tau} 
     \right).
\]
Thus, for arbitrary sections
$ s_{\tau_1} $
and
$ s_{\tau_2} $
of
$ \mathcal{G} $, 
the
$ d $-closed
forms 
$ s^*_{\tau_1} u(\mathcal{G},\nabla) $
and 
$ s^*_{\tau_2} u(\mathcal{G},\nabla) $
differ by a
$ d $-exact
form and hence are cohomologous. 
\end{proof}
\begin{corol}
The form
$ s^* u(\mathcal{G},\nabla) $
is cohomologous to the
\emph{Euler form}
\[
e(\mathcal{G},\nabla) 
   = \frac{1}{(2 \pi)^\frac{r}{2}} \int d \rho \,
     \exp \left[
                 \frac{1}{2} 
                 \left(
                        \rho, \mathcal{R} \rho
                 \right)_{\mathcal{G}^{\vee}}
          \right]
   = \mathrm{Pfaff} \left( 
                           \frac{\mathcal{R}}{2 \pi}
                    \right)
\]
and hence is a representative of the Euler class of
$ \mathcal{G} $.
\end{corol}
\begin{proof}
This follows from proposition \ref{MQ-independent-s} upon choosing $ s $ to be the zero section.
\end{proof}
\begin{remar}
The top Chern class of a complex vector bundle is equal to the Euler class of the underlying real vector bundle.
\end{remar}
\begin{remar}
If
$ s $
intersects the zero section of
$ \mathcal{G} $
transversely, then
$ s^* u(\mathcal{G},\nabla) $
is Poincar\'{e} dual to
$ s^{-1}(0) $,
i.e. 
\begin{equation}
\int_M 
\omega 
\wedge 
s^* u(\mathcal{G},\nabla)
   = \int_{s^{-1}(0)} 
     \omega \, , 
\end{equation}
where
$ \omega \in \Omega^{n-r}(M) $
is
$ d $-closed. 
\end{remar}
\begin{remar}
When
$ n = r $,
integrating
$ s^* u(\mathcal{G},\nabla) $
over
$ M $
yields the Euler number of
$ \mathcal{G} $.
\end{remar}
\section{\label{section:Deformation-pullback-Mathai-Quillen}Deformation of the pullback of a Mathai-Quillen form}

Various analogues of pullbacks of Mathai-Quillen forms were proposed in
\cite{GaravusoSharpe:Analogues}.
Let us now discuss the analogue that is most relevant to this paper, i.e. a deformation of the pullback of a Mathai-Quillen form.

Consider deforming
$ s^* u(\mathcal{G},\nabla) $
to 
\begin{equation}
\label{omega-delta-s}
\omega_{\delta s}(\mathcal{G},\nabla)
   = a_r
     \int d \rho \,
     \exp 
     \left(
          - \mathcal{A}_{\delta s}    
     \right),  
\end{equation}
where
\begin{equation}
\mathcal{A}_{\delta s}
   = \mathcal{A} 
   + \left\langle 
            \rho^{p^{\prime}} f_{p^{\prime}},
            d \phi^i
            \left(
                   \delta s
            \right)_{ip}
            e^p
     \right\rangle_{\mathcal{G}}.
\end{equation}
Here,
$ \mathcal{A} $
is given by
\eqref{mathcal-A}
and
\begin{equation}
\left(
       \delta s
\right)_{ip} 
   \in
   \Gamma
     \left(
            \pi^*
            \mathcal{G}
            \otimes
            \pi^*
            TM
     \right) .         
\end{equation}

The deformation
$ \omega_{\delta s}(\mathcal{G},\nabla) $
and
$ s^* u(\mathcal{G},\nabla) $ 
are special cases of the analogue
$ \omega_{\textrm{K1}} $
of
$ s^* u(\mathcal{G},\nabla) $
proposed in
\cite{GaravusoSharpe:Analogues}.
Briefly,
\[
\omega_{\textrm{K1}}
  \propto
    {\textstyle
    \int
    \left[
           {\textstyle \prod}_{\overline{x}} \,
           d \lambda^{\overline{x}} \,
    \right]
    \left[ 
           \prod_r
           d \chi^r
    \right]}
    \exp
    \left(
         - \mathcal{A_{\textrm{K1}}}
    \right)
  \in
    H^{\, \textrm{rk} \, \mathcal{G}}
    \left(
           M,
           \wedge^{\textrm{rk} \, \mathcal{F}_2}
           \mathcal{F}^{\vee}_2
           \otimes
           \det{\mathcal{G}^{\vee}}
           \otimes
           \det{\mathcal{F}_2}
    \right) ,         
\]
where
$ \mathcal{F}_1 $
and
$ \mathcal{F}_1 $
are holomorphic vector bundles on
$ M $
and
\[
\mathcal{A_{\textrm{K1}}}
  = h^{x \overline{x}}
    s_x
    \overline{s}_{\overline{x}}
  + \chi^{\overline{\imath}}
    \lambda^{\overline{x}} \,
    \overline{D}_{\overline{\imath}} \,
    \overline{s}_{\overline{x}} \,
  + \chi^r
    \lambda^{\gamma}
    \widetilde{F}_{r \gamma}
  + F_{\overline{\imath} r \overline{x} \gamma}
    \chi^{\overline{\imath}}
    \chi^r
    \lambda^{\overline{x}}
    \lambda^{\gamma} \, .      
\]
Here,
$ x $
indexes local coordinates along the fibers of
$ \mathcal{G} $,
$ \gamma $
indexes local coordinates along the fibers of
$ \mathcal{F}_1 $,
$ r $
indexes local coordinates along the fibers of
$ \mathcal{F}^{\vee}_2 $,
and
$ i $
indexes local coordinates on
$ M $.
$ s \in \Gamma{(\mathcal{G})} $.
The map
$ \widetilde{F}: \mathcal{F}_1 \rightarrow \mathcal{F}_2 $
is smooth and surjective.
The curvature term
$ 
  F_{\overline{\imath} r \overline{x} \gamma}
  \chi^{\overline{\imath}}
  \chi^r
  \lambda^{\overline{x}}
  \lambda^{\gamma}
$
is subject to the constraint
\[
\overline{\partial}_{\overline{\imath}}
\widetilde{F}_{r \gamma}
  = h^{x \overline{x}}
    s_x
    F_{\overline{\imath} r \gamma \overline{x}}
  =
  - h^{x \overline{x}}
    s_x 
    F_{\overline{\imath} r \overline{x} \gamma} \,
\]
which is imposed physically by supersymmetry.
Note that this constraint is consistent with the curvature 2-form being
$ \overline{\partial} $-closed
by virtue of the Bianchi identity.
One may show that
\[
\left(
       \overline{D}
     + h^{x \overline{x}}
       s_x
       \frac{\partial}{\partial \lambda^{\overline{x}}}  
\right)
\mathcal{A_{\textrm{K1}}}
  = \chi^i
    \chi^r
    \lambda^{\gamma}
    \left(
           \overline{\partial}_{\overline{\imath}}
           \widetilde{F}_{r \gamma}
         + h^{x \overline{x}}
           s_x 
           F_{\overline{\imath} r \overline{x} \gamma}   
    \right)
  = 0 \, ,  
\]
where
$ \overline{D} = \chi^{\overline{\imath}} \, \overline{\partial}_{\overline{\imath}} $.
It follows that
$ \overline{\partial} \, \omega_{\textrm{K1}} = 0 $.
Let
$ Y \equiv \left\{ s = 0 \right\} \subset M $
and let
$ \mathcal{E}^{\prime} $
be the restriction to
$ Y $
of the kernel of the map
$ \widetilde{F} $.
Then
\[
\int_Y
\mathcal{O}_1
\wedge
\cdots
\wedge
\mathcal{O}_k
  = \int_M
    \widetilde{\mathcal{O}}_1
    \wedge
    \cdots
    \wedge
    \widetilde{\mathcal{O}}_k
    \wedge
    \omega_{\textrm{K1}} \, ,
\]
where
$ 
  \widetilde{\mathcal{O}}_1
  \wedge
  \cdots
  \wedge
  \widetilde{\mathcal{O}}_k
    \in
    H^{\dim{M} - \textrm{rk} \, \mathcal{G}}
    \left(
           M,
           \wedge^{\textrm{rk} \, \mathcal{F}_1 -  \textrm{rk} \, \mathcal{F}_2}
           \mathcal{F}^{\vee}_2
    \right)
$
and
$
  \widetilde{\mathcal{O}}
  \in
  H^{\bullet}
  \left(
         M,
         \wedge^{\bullet}
         \mathcal{F}^{\vee}_1
  \right) 
$
is a lift of
$
  \mathcal{O}
  \in
  H^{\bullet}
  \left(
         Y,
         \wedge^{\bullet}
         \mathcal{\mathcal{E}}^{\prime \vee}
  \right) 
$.
For this reason,
$ \omega_{\textrm{K1}} $
is called in
\cite{GaravusoSharpe:Analogues}
the (first) kernel construction.
See
\cite{GaravusoSharpe:Analogues}
for further details.
One recovers
$ \omega_{\delta s}(\mathcal{G},\nabla) $
and
(when $ \delta s = 0 $)
$ s^* u(\mathcal{G},\nabla) $
in the special case that
$ \mathcal{F}_1 = TM $
and
$ \mathcal{F}_2 = \mathcal{G} $
with the map
$ \widetilde{F}: \mathcal{F}_1 \rightarrow \mathcal{F}_2 $
defined by
\begin{equation}
F_{ip}
  = D_i s_p
  + \left(
           \delta s
    \right)_{ip} \, ,         
\end{equation}  
where
$ s_p $
is a holomorphic section of
$ \mathcal{G} $.
This corresponds to
$ \mathcal{E}^{\prime} $
being a deformation of
$ TY $,
with the deformation determined by
$ \delta s $.
If
$ \delta s = 0 $,
then
$ \mathcal{E}^{\prime} = TY $.
Note that
\[
\overline{\partial}_{\overline{\imath}} \,
F_{ip}
  = \overline{\partial}_{\overline{\imath}}
    \left(
           D_i
           s_p
         + \left(
                  \delta s
           \right)_{ip}
    \right)
  = \left[ \,
           \overline{D}_{\overline{\imath}} \, ,
           D_i
    \right]
    s_p
  = R_{\overline{\imath} i p \overline{p}} \, 
    g^{p \overline{p}}
    s_p \, .     
\] 
\begin{prop}
The form 
$ \omega_{\delta s}(\mathcal{G},\nabla) $
satisfies
\[
\overline{\partial} \,
\omega_{\delta s}(\mathcal{G},\nabla)
   = 0 \, .
\]
\end{prop}
\begin{proof}
For 
$ \mathcal{A} $
given by
\eqref{mathcal-A},
we have that 
$ \overline{D} \mathcal{R} = 0 $
and hence

\begin{align}
\label{Dbar-mathcal-A-equation}
\left(
       \overline{D}        
     + s^{\overline{p}} 
       \frac{\partial}{\partial \rho^{\overline{p}}}
\right)
\mathcal{A} 
  &=
   - \left(
            s_p e^p, \overline{D} s_{\overline{p}} \, e^{\overline{p}}
     \right)_{\mathcal{G}}
   + \left\langle
            \mathcal{R} s_{p^{\prime}} e^{p^{\prime}}, \rho^p f_p  
     \right\rangle_{\mathcal{G}}
   - \frac{1}{2}
     \left(
            \rho, \overline{D} \mathcal{R} \rho
     \right)_{\mathcal{G}^{\vee}}
\nonumber
\\[1ex]
  &\phantom{=}
   + \left(
            s_p e^p, \overline{D} s_{\overline{p}} \, e^{\overline{p}}
     \right)_{\mathcal{G}}
   - \left\langle
            \mathcal{R} s_{p^{\prime}} e^{p^{\prime}}, \rho^p f_p 
     \right\rangle_{\mathcal{G}}
\nonumber
\\[2ex]
  &= 0 \, .                             
\end{align}
It follows that
\begin{equation}
\left(
       \overline{D}        
     + s^{\overline{p}} 
       \frac{\partial}{\partial \rho^{\overline{p}}}
\right)
\mathcal{A}_{\delta s}
   = \left(
            \overline{D}        
          + s^{\overline{p}} 
            \frac{\partial}{\partial \rho^{\overline{p}}}   
      \right)
      \left[
             \mathcal{A}
           + \left\langle 
                   \rho^{p^{\prime}} f_{p^{\prime}},
                   d \phi^i
                   \left(
                          \delta s
                   \right)_{ip}
                   e^p
              \right\rangle_{\mathcal{G}}             
      \right]
   = 0 \, .             
\end{equation}
Using this result, we obtain
\[
\begin{aligned}
\overline{\partial} \,
\omega_{\delta s}(\mathcal{G},\nabla)
  &= a_r \,
     \overline{\partial}
     \int d \rho \,
     \exp
     \left(
          - \mathcal{A}_{\delta s}
     \right)  
\nonumber
\\[1ex]
  &= a_r
     \int d \rho \,
     \overline{D}
     \exp
     \left(
          - \mathcal{A}_{\delta s}
     \right)
\nonumber
\\[1ex]
  &= a_r
     \int d \rho \,
     \left(
            \overline{D}        
          + s^{\overline{p}} 
            \frac{\partial}{\partial \rho^{\overline{p}}}  
     \right) 
     \exp
     \left(
          - \mathcal{A}_{\delta s}
     \right)
\nonumber
\\[1ex]
  &= a_r
     \int d \rho \,
     \left[
          - \left(
                   \overline{D}        
                 + s^{\overline{p}} 
                   \frac{\partial}{\partial \rho^{\overline{p}}}            
            \right)
            \mathcal{A}_{\delta s} 
     \right]
     \exp 
     \left(
          - \mathcal{A}_{\delta s}
     \right)
\nonumber     
\\[1ex]
  &= 0 \, .
\end{aligned}
\]
\end{proof}
\begin{prop}
\label{deformed-MQ-s-independent-antiholomorphic}
The
$ \overline{\partial} $-cohomology
class of
$ \omega_{\delta s}(\mathcal{G},\nabla) $
is unchanged by antiholomorphic deformations of
$ s $. 
\end{prop}
\begin{proof}
Let
$ s_{\alpha} = s + \alpha \, s^{\prime}_{\overline{p}} \, e^{\overline{p}} $
be an affine one parameter family of sections of
$ \mathcal{G} $
and let  
\[
\mathcal{A}_{\delta s,\alpha}
   = \frac{1}{2}
     \left(
            s_{\alpha}, s_{\alpha}
     \right)_{\mathcal{G}}
   + \left(
            \nabla s_{\alpha}, \rho
     \right)_{\mathcal{G}}
   + \frac{1}{2}
     \left(
            \rho, \mathcal{R} \rho
     \right)_{\mathcal{G}^{\vee}}
   + \left\langle
            \rho^{p^{\prime}} f_{p^{\prime}},
            d \phi^i
            \left(
                   \delta s
            \right)_{ip}
            e^p
     \right\rangle_{\mathcal{G}} .
\]
Then
\[
\begin{aligned}
\frac{d}{d \alpha}
\omega_{\delta s,\alpha}(\mathcal{G},\nabla)
   &= a_r \, 
      \frac{d}{d \alpha}
      \int d \rho \,
      \exp
      \left(
           - \mathcal{A}_{\delta s,\alpha}
      \right)      
\nonumber
\\[1ex]
  &= 
   - a_r
     \int d \rho \,
     \left[
            \left(
                   s^{\prime}_{\overline{p}} \, e^{\overline{p}},
                   s_p e^p
            \right)_{\mathcal{G}}
          + \left\langle
                   \rho^{\overline{p}} f_{\overline{p}},
                   \overline{D} s^{\prime}_{\overline{p}^{\prime}} \, 
                   e^{\overline{p}^{\prime}}
            \right\rangle_{\mathcal{G}}                
     \right]
     \exp
     \left(
          - \mathcal{A}_{\delta s,\alpha}
     \right)
\nonumber
\\[1ex]
  &= 
   - a_r
     \int d \rho \,
     \left[
            \left(
                   \overline{D}        
                 + s^{\overline{p}} 
                   \frac{\partial}{\partial \rho^{\overline{p}}}  
            \right)
            \left\langle
                   \rho^{\overline{p}} f_{\overline{p}},
                   s^{\prime}_{\overline{p}^{\prime}} \, e^{\overline{p}^{\prime}}
            \right\rangle_{\mathcal{G}}                
     \right]
     \exp
     \left(
          - \mathcal{A}_{\delta s,\alpha}
     \right)
\nonumber
\\[1ex]
  &= 
   - a_r
     \int d \rho \,
     \left(
            \overline{D}        
          + s^{\overline{p}} 
            \frac{\partial}{\partial \rho^{\overline{p}}}       
     \right)
     \left[
            \left\langle
                   \rho^{\overline{p}} f_{\overline{p}},
                   s^{\prime}_{\overline{p}^{\prime}} \, e^{\overline{p}^{\prime}}
            \right\rangle_{\mathcal{G}}               
            \exp
            \left(
                 - \mathcal{A}_{\delta s, \alpha}
            \right)
     \right]
\nonumber
\\[1ex]
  &= 
   - a_r \,
     \overline{\partial}
     \int d \rho \,
     \left\langle  
            \rho^{\overline{p}} f_{\overline{p}},
            s^{\prime}_{\overline{p}^{\prime}} \, e^{\overline{p}^{\prime}}            
     \right\rangle_{\mathcal{G}} \,                
     \exp
     \left(
          - \mathcal{A}_{\delta s,\alpha}
     \right).                                          
\end{aligned}
\]
It follows that
\[
  \omega_{\delta s,\alpha_2}(\mathcal{G},\nabla)
- \omega_{\delta s,\alpha_1}(\mathcal{G},\nabla)
= 
- a_r \,
  \overline{\partial}
  \int^{\alpha_2}_{\alpha_1} d \alpha
  \int d \rho \,
  \left\langle
         \rho^{\overline{p}} f_{\overline{p}},
         s^{\prime}_{\overline{p}^{\prime}} \, e^{\overline{p}^{\prime}}
  \right\rangle_{\mathcal{G}}                
  \exp
  \left(
       - \mathcal{A}_{\delta s,\alpha}
  \right),    
\]
which establishes that the 
$ \overline{\partial} $-cohomology
class of 
$ \omega_{\delta s}(\mathcal{G},\nabla) $
is unchanged by antiholomorphic deformations of 
$ s $.
\end{proof}
\begin{remar}
\label{deformed-MQ-s-dependent-tangent-bundle}
The
$ \overline{\partial} $-cohomology
class of 
$ \omega_{\delta s}(\mathcal{G},\nabla) $ 
does seem to depend on the choice of the 
$ \left( \delta s\right)_{ip} $,
at least naively.
Let 
$ 
\left(
       \delta s
\right)_{ip,\gamma} 
   = \left( 
            \delta s
      \right)_{ip} 
   + \gamma 
     \left(
            \delta s
      \right)^{\prime}_{ip}
$
and
$
\mathcal{A}_{\delta s,\gamma}
  = \mathcal{A} 
  + \left\langle
           \rho^{p^{\prime}} f_{p^{\prime}},
           d \phi^i
           \left(
                   \delta s
           \right)^{\gamma}_{ip}
           e^p
     \right\rangle_{\mathcal{G}} .
$
Then
\[
\begin{aligned}
\frac{d}{d \gamma} 
\omega_{\delta s,\gamma}(\mathcal{G},\nabla)
  &= a_r \,
     \frac{d}{d \gamma}
     \int d \rho \,
     \exp
      \left(
           - \mathcal{A}_{\delta s,\gamma}  
     \right)
\\[1ex]
  &= 
   - a_r
     \int d \rho \,
     \left\langle
            \rho^{p^{\prime}} f_{p^{\prime}},
            d \phi^i 
            \left(
                   \delta s 
            \right)^{\prime}_{ip}
            e^p
     \right\rangle_{\mathcal{G}}
     \exp
     \left(
           - \mathcal{A}_{\delta s,\gamma}  
     \right).          
\end{aligned}
\]
It follows that
\[
  \omega_{\delta s,\gamma_2}(\mathcal{G},\nabla) 
- \omega_{\delta s,\gamma_1}(\mathcal{G},\nabla)
   = 
   - a_r \int^{\gamma_2}_{\gamma_1} d \gamma
     \int d \rho \,
     \left\langle
            \rho^{p^{\prime}} f_{p^{\prime}},
            d \phi^i 
            \left(
                   \delta s
             \right)^{\prime}_{ip}
            e^p
     \right\rangle_{\mathcal{G}}
     \exp
     \left(
          - \mathcal{A}_{\delta s,\gamma}  
     \right),          
\]
which is at least not obviously 
$ \overline{\partial} $-exact.
The physical meaning of this result is commented on in
\cite{GaravusoSharpe:Analogues}.
\end{remar}

\end{appendix}

\end{document}